\documentclass[aps,prd,superscriptaddress,showpacs,A4paper,preprintnumbers]{revtex4}
\usepackage{graphicx,color}
\usepackage{bm}
\newcommand{\be}{\begin{equation}}
\newcommand{\ee}{\end{equation}}
\newcommand{\bea}{\begin{eqnarray}}
\newcommand{\eea}{\end{eqnarray}}

\newcommand{\bfk}{\mbox{\boldmath $k$}}

\newcommand{\bfp}{\mbox{\boldmath $p$}}

\newcommand{\pup}{p^\uparrow}

\def\lsim{\mathrel{\rlap{\lower4pt\hbox{\hskip1pt$\sim$}}\raise1pt\hbox{$<$}}}
\def\gsim{\mathrel{\rlap{\lower4pt\hbox{\hskip1pt$\sim$}}\raise1pt\hbox{$>$}}}
\def\nostrocostruttino#1\over#2{\mathrel{\mathop{\kern 0pt \rlap
{\hbox{$#1$}}} \hbox{\kern-.135em $#2$}}}

\begin{document}

\title{Towards a first estimate of the gluon Sivers function\\ from $A_N$ data in $pp$ collisions at RHIC}

\author{U.~D'Alesio}
\affiliation{Dipartimento di Fisica, Universit\`a di Cagliari, Cittadella Universitaria, I-09042 Monserrato (CA), Italy}
\affiliation{INFN, Sezione di Cagliari, C.P.~170, I-09042 Monserrato (CA), Italy}
\author{F.~Murgia}
\affiliation{INFN, Sezione di Cagliari, C.P.~170, I-09042 Monserrato (CA), Italy}
\author{C.~Pisano}
\affiliation{Department of Physics, University of Antwerp, Groenenborgerlaan 171, 2020 Antwerp, Belgium}
\date{\today}

\begin{abstract}
Within a generalized parton model approach, with inclusion of spin and intrinsic transverse momentum effects,
we show how the latest, highly precise, midrapidity data on the transverse single spin asymmetry measured in $pp\to\pi^0\, X$ by the PHENIX Collaboration at RHIC~\cite{Adare:2013ekj}, can be used to get a first estimate on the still poorly known gluon Sivers distribution. To this end we also adopt the present information on the quark Sivers functions, as extracted from semi-inclusive deeply inelastic scattering data.
This analysis updates a previous study by some of us where a first bound on this distribution was obtained~\cite{Anselmino:2006yq}.
\end{abstract}

\pacs{13.88.+e, 12.38.Bx, 13.85.Ni}

\maketitle

\section{\label{1}Introduction and formalism}

The study of the 3-dimensional nucleon structure is, nowadays, one of the most interesting and challenging topics in hadron physics. In the last decade it has become clear that even in high-energy processes a one-dimensional picture of the nucleon in terms of collinear parton distribution functions (PDFs) is not always satisfactory and a more complete description involving also transverse degrees of freedom, both in spin and momentum, is necessary. To this aim a new class of transverse momentum dependent parton distributions (TMD-PDFs) and fragmentation functions (TMD-FFs), shortly referred to as TMDs, has been introduced~(see e.g.~Refs.~\cite{D'Alesio:2007jt,Barone:2010zz} for review).

Many transverse spin phenomena in hard processes, like the well-known transverse single spin asymmetries (SSAs) observed in inclusive hadron production in hadron-hadron collisions as well as the more recent azimuthal asymmetries measured in semi-inclusive deeply inelastic scattering (SIDIS) processes with a transversely polarized target, have challenged the full theoretical understanding of QCD.

At present, two main theoretical schemes have been formulated to deal with these transverse spin asymmetries: one, originally proposed in Refs.~\cite{Efremov:1981sh,Efremov:1984ip,Qiu:1991pp,Qiu:1998ia,Kanazawa:2000hz} and phenomenologically developed in Refs.~\cite{Kouvaris:2006zy,Eguchi:2006mc,Koike:2009ge,Kanazawa:2014dca}, is based on collinear higher-twist parton correlators. This formalism has been proved to be valid for processes where only one hard scale is present, like the transverse momentum of the final particles inclusively produced in $pp$ collisions. A second approach, based on TMD factorization theorems, was shown to be valid for processes characterized by two energy scales: a hard one, like the virtuality of the exchanged boson in SIDIS, Drell-Yan processes (DY) or $e^+e^-$ annihilation and a soft one, comparable to $\Lambda_{\rm QCD}$, like the transverse momentum of the final hadron in
SIDIS, or of the lepton pair in DY, or the transverse momentum imbalance in hadron-pair production in $e^+e^-$ collisions~\cite{Ji:2004wu,Ji:2004xq, Bacchetta:2008xw, Collins:2011zzd, GarciaEchevarria:2011rb}. Expressions of azimuthal asymmetries in terms of TMDs for such processes can be found in Refs.~\cite{Bacchetta:2006tn,Anselmino:2011ch} (SIDIS), Refs.~\cite{Boer:1999mm,Anselmino:2002pd,Arnold:2008kf} (DY), and Refs.~\cite{Boer:1997mf,Anselmino:2007fs} ($e^+e^-$).

The two approaches are related in the existing common region of validity~\cite{Ji:2006ub,Ji:2006vf,Boer:2003cm, Koike:2007dg,Yuan:2009dw}, although a formal proof of factorization in single-particle production still lacks for the TMD approach (see e.g.~Ref.~\cite{Rogers:2013zha}).

Concerning the QCD scale evolution of TMDs much progress has been also done in the last years~\cite{Collins:2011zzd, Aybat:2011zv, Aybat:2011ta, Anselmino:2012aa, Boer:2013zca, Sun:2013hua, Aidala:2014hva, Echevarria:2014xaa, Collins:2014jpa, D'Alesio:2014vja,Echevarria:2015uaa}, although different phenomenological attempts and schemes have been proposed, and a univocal unambiguous treatment is still missing.

Much phenomenological information has been by now collected on quark TMDs; in particular the Sivers distribution~\cite{Sivers:1989cc, Sivers:1990fh} and the Collins fragmentation function~\cite{Collins:1992kk} have been extracted from SIDIS and $e^+e^-$ data by different groups~\cite{Anselmino:2005ea, Vogelsang:2005cs, Collins:2005ie,Efremov:2006qm, Anselmino:2007fs, Anselmino:2008jk, Anselmino:2008sga, Anselmino:2012rq, Anselmino:2013vqa, Anselmino:2013rya}. Even if to a lesser extent some information on the Boer-Mulders function~\cite{Boer:1997nt} has been also gathered~\cite{Gamberg:2007wm,Zhang:2008nu, Lu:2009ip,Barone:2009hw, Pasquini:2010af,Barone:2010gk,Musch:2011er}. On the other hand, up to now, very little is known on gluon TMDs.

The gluon Sivers function (GSF), for instance, is constrained by a trivial positivity bound (given by two times the unpolarized TMD gluon distribution), which however is very loose and of little usefulness.
A more important theoretical constraint comes from the so-called Burkardt sum rule (BSR)~\cite{Burkardt:2004ur}.
It states, in a non-trivial way due to the presence of QCD color-gauge links, the vanishing of the total transverse momentum of all unpolarized partons inside a transversely polarized proton. Fits to the Sivers asymmetry for SIDIS data in the TMD approach~\cite{Anselmino:2005ea,Anselmino:2008sga} almost fulfil, within uncertainties, the BSR, leaving little space for a gluon contribution.
In the large-$N_c$ limit of QCD the GSF should be suppressed by a factor $1/N_c$ w.r.t.~the valence quark Sivers
distributions, at not too small Bjorken-$x$ values ($x\sim 1/N_c$) (see Ref.~\cite{Efremov:2004tp} and references
therein).
The COMPASS Collaboration at CERN is currently studying the Sivers asymmetry in the production of high-$p_T$ hadron pairs in muon scattering off polarized proton and deuteron targets~\cite{Szabelski:2015zzz}.
This process should be dominated by the photon-gluon fusion mechanism and therefore allows to get information on the GSF. First results gave an asymmetry compatible with zero for deuteron target at $\langle x_G \rangle = 0.13$.
This fact, together with additional theoretical considerations, led Brodsky and Gardner~\cite{Brodsky:2006ha}
to state that the gluon contribution to parton orbital angular momentum (and the GSF) should be negligible.
However, very recent preliminary measurements of the same observable for proton target give a negative gluon Sivers asymmetry, $-0.26\pm 0.09\pm 0.08$ at $\langle x_G \rangle = 0.15$~\cite{Szabelski:2015zzz}. This value even if $3\sigma$ below zero is still compatible with the deuteron result.

{}From the phenomenological point of view, it has been suggested to study the role of the GSF in polarized
proton-proton collisions in several processes: SSAs in inclusive photon production in the large negative $x_F$ region (measured w.r.t.~the polarized proton)~\cite{Schmidt:2005gv}; back-to-back azimuthal correlations in two-jet
production~\cite{Boer:2003tx}; SSAs in inclusive $D$ meson production at RHIC~\cite{Anselmino:2004nk};
SSAs in $J/\psi$ electroproduction with transversely polarized electron and proton beams~\cite{Godbole:2014tha}. A detailed and updated discussion on the gluon Sivers function and additional references can be found in Ref.~\cite{Boer:2015vga}.

Besides the gluon Sivers function, the role of linearly polarized gluons inside (un)polarized protons in inclusive processes in proton-proton collisions has been also actively investigated in recent years, e.g.~in pion-jet production~\cite{D'Alesio:2010am}, heavy quark and jet-pair production at electron-ion or hadron colliders~\cite{Boer:2010zf,Pisano:2013cya}, and Higgs production at the
LHC~\cite{Boer:2011kf,Boer:2014tka,Echevarria:2015uaa}.

Assuming the validity of the TMD formalism for a single-scale process we show here how the analysis of highly precise  midrapidity data in single polarized $pp\to \pi X$ processes could strongly constrain the gluon Sivers function. As shown in a series of papers~\cite{Anselmino:1994tv,D'Alesio:2004up,Anselmino:2005sh,Anselmino:2011ch, Anselmino:2012rq,Anselmino:2013rya} this phenomenological approach, nowadays known as generalized parton model (GPM) is able to describe fairly well many features of several available data for such a process and it is worth to be further investigated. Even if not supported, as already said above, by a formal proof of factorization in terms of TMDs, the study of these processes can be very useful also in clarifying the role of process dependence and factorization breaking effects.

It has been already shown that, within a TMD scheme, due to strong partonic azimuthal phase cancellations, the backward hemisphere can be of little use to get information on polarized TMDs, since all effects are almost washed out~\cite{Anselmino:2005sh}. On the other hand, as shown by some of us in Ref.~\cite{Anselmino:2006yq}, a study of midrapidity $A_N$ data at high energy can be used to constrain the gluon Sivers function. Here we will upgrade this result by using more recent information both on the phenomenological and the experimental side.

Indeed we have now at our disposal phenomenological extractions of the quark Sivers functions from SIDIS processes (not available at that time), one of them including also the sea quark contributions~\cite{Anselmino:2005ea,Anselmino:2008sga}. At the same time new and highly precise data from the PHENIX Collaboration at RHIC have been made available~\cite{Adare:2013ekj}. For these reasons we believe that such a reanalysis is timely.

As said, the issue of QCD evolution of TMDs, strongly related to factorization, is still an open question for such single scale processes. Concerning its potential role in the following analysis, we believe that the
relatively modest range of pion transverse momentum involved (at least for the more precise data which can significantly constrain the GSF) prevents it to be effective for the asymmetries.
Therefore, in the sequel we will keep including QCD evolution only in the collinear factorized component of the involved TMDs (see also below for more details).

Although in the TMD approach several terms may in principle contribute to the single spin asymmetry $A_N(p^\uparrow p\to \pi\,X)\equiv (d\sigma^\uparrow-d\sigma^\downarrow)/(d\sigma^\uparrow+d\sigma^\downarrow)$, in the kinematical regime of Ref.~\cite{Adare:2013ekj}, as extensively discussed in~Ref.~\cite{Anselmino:2006yq}, $A_N$ is largely dominated by the Sivers effect alone, and its numerator is given by (for details see Refs.~\cite{D'Alesio:2004up,Anselmino:2005sh})
\bea \!\!\!\!\! && \frac{E_\pi \, d\sigma^\uparrow}{d^3\bfp_\pi} -
\frac{E_\pi \, d\sigma^\downarrow}{d^3\bfp_\pi} \simeq \sum_{a,b,c,d} \int \frac{dx_a \, dx_b \, dz}{\pi \, x_a \, x_b \, z^2 \, s} \; d^2\bfk_{\perp a} \, d^2\bfk_{\perp b} \,
\,d^3\bfk_{\perp \pi} \, \delta(\bfk_{\perp \pi}\cdot \hat{\bfp}_c) \, J(k_{\perp \pi})
\nonumber \\ \!\!\!\!\! &\times& \Delta \hat f_{a/\pup}(x_a, \bfk_{\perp a})
\> f_{b/p}(x_b, k_{\perp b})\,\hat s^2 \>\frac{d\hat\sigma^{ab \to cd}}
{d\hat t}(x_a,x_b, \hat s, \hat t, \hat u) \> \delta(\hat s + \hat t + \hat u) \>  D_{\pi/c}(z, k_{\perp \pi}) \>,
\label{sivgen}
\eea
where ($M_p$ denotes the proton mass)
\bea
\Delta \hat f_{a/\pup}\,(x_a, \bfk_{\perp a}) &\equiv&
\hat f_{a/\pup}\,(x_a, \bfk_{\perp a}) - \hat f_{a/p^\downarrow}\,
(x_a, \bfk_{\perp a})\nonumber \\
\label{defsiv}
&=& \Delta^N f_{a/\pup}\,(x_a, k_{\perp a}) \> \cos\phi_a\nonumber \\
&=&  -2 \, \frac{k_{\perp a}}{M_p} \, f_{1T}^{\perp a} (x_a, k_{\perp a}) \>
\cos\phi_a \>.
\eea
$\Delta^N f_{a/\pup}(x_a, k_{\perp a})$
[or $f_{1T}^{\perp a} (x_a, k_{\perp a})]$ is referred to as the Sivers distribution function of parton $a$ inside a transversely polarized proton~\cite{Bacchetta:2004jz}. $\phi_a$ is the azimuthal angle of the intrinsic transverse momentum $\bfk_{\perp a}$ of parton $a$. For details and a full explanation of the notations in Eq.~(\ref{sivgen})
see Ref.~\cite{Anselmino:2005sh}. It suffices to notice here that $J(k_{\perp\pi})$ is a kinematical factor, which at ${\cal O}(k_{\perp\pi}/E_{\pi})$ equals 1 and $d\hat\sigma^{ab \to cd}/d\hat t$ is the partonic differential cross section for the subprocess $ab\to cd$.

Notice that the parton $a$ inside the polarized proton can be a quark (or an antiquark) and a gluon, that is the Sivers contribution to the asymmetry can be expressed as a sum of two terms,
\be
A_N = A_N^{\rm quark} + A_N^{\rm gluon}\,,
\ee
that cannot be disentangled in this process.

For this reason in the following analysis we will take into account all available information on the quark Sivers functions.
In particular we consider two extractions. The reason for this is twofold: from one side,
in the first extraction (SIDIS1 in the following)~\cite{Anselmino:2005ea} only $u$ and $d$ flavours were considered,
while in the second one (SIDIS2)~\cite{Anselmino:2008sga}  also the sea quark contributions were accounted for; secondly, and somehow more relevant, in SIDIS1 the set of fragmentation functions of Kretzer~\cite{Kretzer:2000yf} was adopted, while in SIDIS2 the set of de Florian, Sassot and Stratmann (DSS)~\cite{deFlorian:2007aj} was considered, which provides a much more important and very different leading order gluon fragmentation.
This aspect, as shown in the following, could play a non negligible role in the present study.

\section{Analysis and results}

The presently available data on $A_N(p^\uparrow p\to \pi^0\,X)$ by the PHENIX Collaboration~\cite{Adare:2013ekj} are extremely precise, of the order of per mil, and with tiny errors, in particular in the region of moderate $P_T$, where the gluon initiated processes dominate. In fact, both their central values and the error bars are at least one order of magnitude smaller than the data analysed in Ref.~\cite{Anselmino:2006yq}. For this reason, while in that work a first, very conservative, upper bound on the gluon Sivers function was presented without entering into a more detailed analysis, here we want to present a more careful study aiming at a first tentative estimate of the GSF within a TMD scheme.

As stated above, in the kinematical region considered only the Sivers effect can play a relevant role, being all the other effects suppressed by strong azimuthal phase cancellations, as discussed in Ref.~\cite{Anselmino:2006yq}.
If one adopts the more detailed SIDIS2 parameterizations of Ref.~\cite{Anselmino:2008sga}, where also the sea quark Sivers functions were considered, one would find a contribution to $A_N$ compatible with zero. Taking into account that the data~\cite{Adare:2013ekj} are also almost compatible with zero, one would conclude that there is no room for the gluon Sivers effect.

In the spirit of Ref.~\cite{Anselmino:2006yq} we adopt again a conservative attitude and investigate to what extent, taking into account the uncertainty on the quark Sivers distributions together with the small errors on the data, a gluon Sivers function could still play a role.

Even if the small number of data points available (ten in this case) does not allow a full statistical analysis, namely a fit, we still try to substantiate our study adopting a commonly used statistical criterium. We then define a proper $\chi^2$ function and, using information available both on the experimental and the phenomenological side, we extract a parametrization of the gluon Sivers function by minimizing it.
Since we aim at extracting the contribution to the Sivers effect from gluons, using the available phenomenological information on the corresponding contribution from quarks with its uncertainty, we define
\be
\label{chi2}
\chi^2 = \sum \frac{(A_N^{\rm gluon} + A_N^{\rm quark} - A_N^{\rm exp})^2}{\sigma_{\rm exp}^2 + \sigma_{\rm quark}^2}\,,
\ee
where the sum runs over the data points, $\sigma_{\rm exp}$ is the experimental error on $A_N^{\rm exp}$ and $\sigma_{\rm quark}$ the \emph{estimated} theoretical uncertainty on the quark contribution $A_N^{\rm quark}$, considered as a known quantity.
In such a way we can constrain the gluon contribution taking into account both the theoretical (even if  partially) and the experimental uncertainties.

Concerning the gluon Sivers function we adopt a somehow standard factorized functional form,
analogous to the quark case~\cite{Anselmino:2005ea,Anselmino:2008sga}, namely:
\be
\Delta^N\! f_{g/p^\uparrow}(x,k_\perp) = 2 \, {\cal N}_g(x)\,f_{g/p}(x)\,
h(k_\perp)\,\frac{e^{-k_\perp^2/\langle k_\perp^2 \rangle}}
{\pi \langle k_\perp^2 \rangle}\,,
\label{eq:siv-par}
\ee
where $f_{g/p}(x)$ is the standard unpolarized collinear gluon distribution,
\be
{\cal N}_g(x) = N_g x^{\alpha}(1-x)^{\beta}\,
\frac{(\alpha+\beta)^{(\alpha+\beta)}}
{\alpha^{\alpha}\beta^{\beta}}\,,
\label{eq:nq-coll}
\ee
with $|N_g|\leq 1$, and
\be
h(k_\perp) = \sqrt{2e}\,\frac{k_\perp}{M'}\,e^{-k_\perp^2/M'^2}\,.
\label{eq:h-siv}
\ee
With these choices, assuming that the unpolarized TMD gluon distribution is given by
\be
f_{g/p}(x,k_\perp) = f_{g/p}(x) \frac{e^{-k_\perp^2/\langle k_\perp^2 \rangle}}{\pi \langle k_\perp^2 \rangle}\,,
\ee
the Sivers function automatically fulfils its proper positivity bound for any $(x,k_\perp)$ values. Consistently, for the unpolarized TMD fragmentation function (for a parton $c$) we use
\be
D_{\pi/c}(z,k_{\perp\pi}) = D_{\pi/c}(z)\,
\frac{e^{-k_{\perp\pi}^2/\langle k_{\perp\pi}^2 \rangle}}
{\pi \langle k_{\perp\pi}^2 \rangle}
\quad\quad\quad \langle k_{\perp\pi}^2\rangle = 0.20\, {\rm GeV}^2 \>.
\ee

In the following, for the Gaussian width of the unpolarized TMD gluon distribution we use the same value
as for the quark distribution, that is $\langle k_\perp^2\rangle = 0.25$ GeV$^2$~\cite{Anselmino:2005ea}.
Moreover, we define the parameter
\be
\label{rho}
\rho = \frac{M'^2}{\langle k_\perp^2 \rangle +M'^2}\,,
\label{eq:rho}
\ee
so that the $k_\perp$-dependent part of the Sivers function becomes
\be
\label{hkrho}
h(k_\perp)\, \frac{e^{-k_\perp^2/\langle k_\perp^2 \rangle}}{\pi \langle k_\perp^2 \rangle} =
\frac{\sqrt{2e}}{\pi}
\,\frac{k_\perp}{M'}\,e^{-k_\perp^2/M'^2} \, \frac{e^{-k_\perp^2/\langle k_\perp^2 \rangle}}{\langle k_\perp^2 \rangle} = \frac{\sqrt{2e}}{\pi} \,\sqrt{\frac{1-\rho}{\rho}}\,k_\perp \, \frac{e^{-k_\perp^2/\rho \langle k_\perp^2 \rangle}}{\langle k_\perp^2 \rangle^{3/2}}\,.
\ee
{}From Eq.~(\ref{eq:rho}) it is clear how the range of variation of the parameter $\rho$ is 0-1.
In the following analysis we will also consider $\rho$ as a free parameter, another improvement w.r.t.~Ref.~\cite{Anselmino:2006yq}, where a fixed value of $\rho$ was adopted in order to maximize the gluon Sivers effect.

We will then minimize the $\chi^2$ function defined in Eq.~(\ref{chi2}) in terms of the following four parameters: $N_g$, $\alpha$, $\beta$ entering Eq.~(\ref{eq:nq-coll}) and $\rho$ in Eqs.~(\ref{eq:rho}),~(\ref{hkrho}).
Since the integration in Eq.~(\ref{sivgen}) is over an eight-dimensional phase space, the fit procedure over
the continuous parameter phase space would be quite CPU-time consuming.
Therefore, we scan the 4-dimensional parameter space over a discrete grid of values, fine enough for our purposes. More precisely, we consider the following ranges: $-1\le N_g \le 1$ (step value of 0.05), $0 \le \alpha, \beta \le 4$ (step value of 0.2), while for the $\rho$ parameter we consider five representative values: 2/3 (as adopted in Ref.~\cite{Anselmino:2006yq} maximizing the effect of the GSF), the same value as for the quark Sivers function and three more values, 0.2, 0.3, and 0.8 (lower or larger values would spoil the description of data).

As stated in the introduction, we will consider the results based on the more recent DSS-SIDIS2 parameterization~\cite{Anselmino:2008sga} as well as those obtained adopting the KRE-SIDIS1 set~\cite{Anselmino:2005ea}, being quite representative extractions of the quark Sivers functions. In both cases, for consistency, we will adopt the GRV98-LO set~\cite{Gluck:1998xa} for the unpolarized parton distributions.

As a first step we checked that the unpolarized cross sections in the same kinematical regime, that is $\sqrt s = 200$ GeV and central rapidity, can be reproduced adopting the TMD distributions and fragmentation functions discussed above. This is an important issue since these quantities enter as denominators in $A_N$.

After that, we performed our $\chi^2$ minimization over the discretized parameter phase-space.
The best (total) $\chi^2$ value obtained is $\chi^2_{\rm min} = 1.93$ for the SIDIS2 set and 1.86 for the SIDIS1 set. Interestingly, the corresponding best value of $\rho$, in both cases, is equal to the corresponding one obtained for the quark Sivers function.

Notice that since the parameters are quite correlated among them, many sets in the explored grid give $\chi^2$'s very close to the minimum value and therefore comparable estimates (see below for a discussion on the uncertainties).
An important remark is that about half of the best $\chi^2$ value comes from the largest-$P_T$ data point,
which has a very large error bar and is less sensitive to the gluon distributions (largest $x$).
Exclusion of this point would give a total minimum $\chi^2$ of about 1.

For completeness, even if this is not the main aim of our study, and taking them with a grain of salt (see previous comments), we give the best-fit parameter sets:
\bea
N_g = 0.05 & \hspace*{1cm} \alpha =0.8 & \hspace*{1cm} \beta =1.4 \hspace*{1.5cm} \rho = 0.576 \label{eq:best-sidis2} \hspace*{2cm} {\rm (SIDIS2)}\\
N_g = 0.65 & \hspace*{1cm} \alpha =2.8 & \hspace*{1cm} \beta =2.8 \hspace*{1.5cm} \rho = 0.687 \label{eq:best-sidis1} \hspace*{2cm} {\rm (SIDIS1)}\,.
\eea

Bearing in mind the caution raised above, it is nevertheless interesting to note that in both cases the gluon Sivers function turns out to be positive. This is another improvement w.r.t.~the previous study~\cite{Anselmino:2006yq} where no information on its sign could be extracted.

In the spirit of being conservative and including potential sources of uncertainties, we will consider also two possible uncertainty bands, generated respectively by the envelope of the $A_N$ values obtained adopting all parameter sets in the parameter-space grid leading to an increase in the $\chi^2$ value of 2\% and 10\% w.r.t.~the $\chi^2_{\rm min}$, that is $\Delta\chi^2 = (2- 10\%)\, \chi^2_{\rm min}$. We notice here that a tolerance of 5\% would give results very similar to the 10\% uncertainty band.
As stated above, given the limited number of experimental data, we cannot claim to have a statistically significant best fit. Therefore, it would not make sense defining and showing statistical error bands. On the other hand, it is useful to quantify the level of accuracy in the description of the data and the corresponding gluon Sivers function when the $\chi^2$ varies within these ranges.

\begin{figure}
\includegraphics[width=8cm,angle=0]{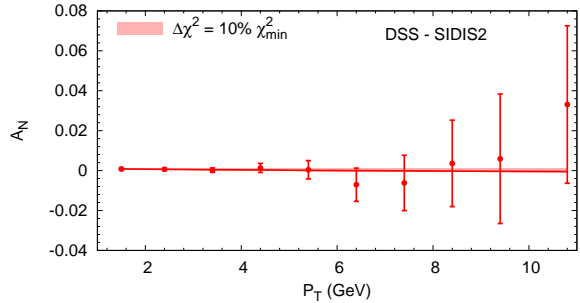}
\caption{Best estimate of the SSA $A_N$, red solid line, compared with PHENIX data~\cite{Adare:2013ekj} at $\sqrt{s}=200$ GeV and at midrapidity, as a function of $P_T$ and adopting the SIDIS2 extraction for the quark Sivers functions~\cite{Anselmino:2008sga}. The red band represents a tolerance of 10\% in $\chi^2$ (see text for details).
 \label{an-dss}}
\end{figure}

\begin{figure}
\includegraphics[width=8cm,angle=0]{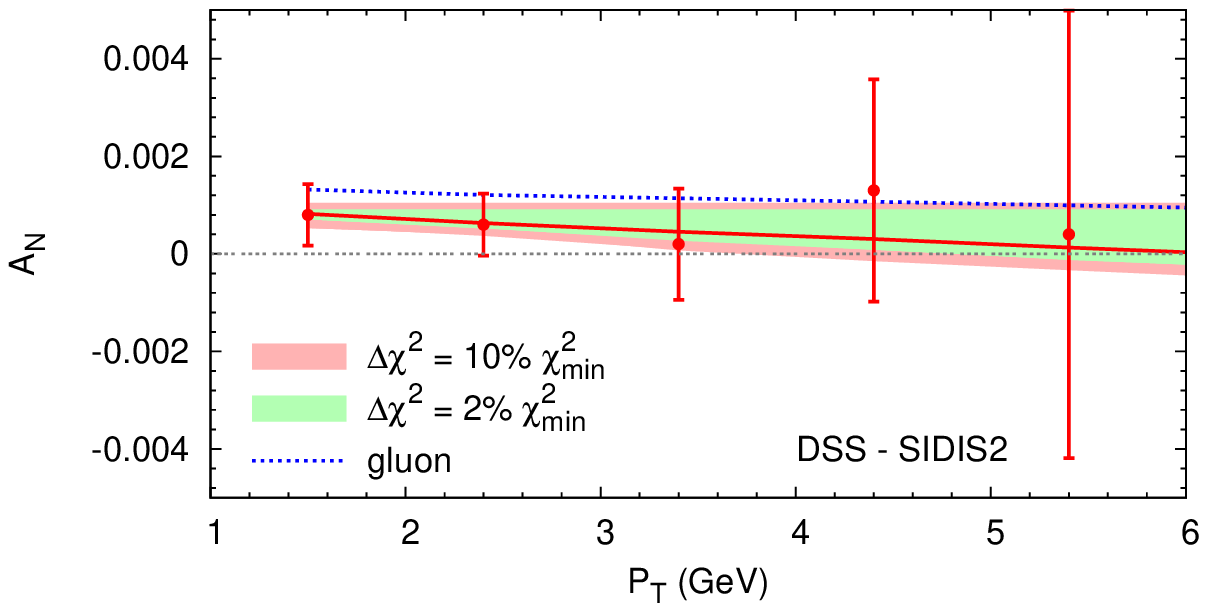}
\includegraphics[width=8cm,angle=0]{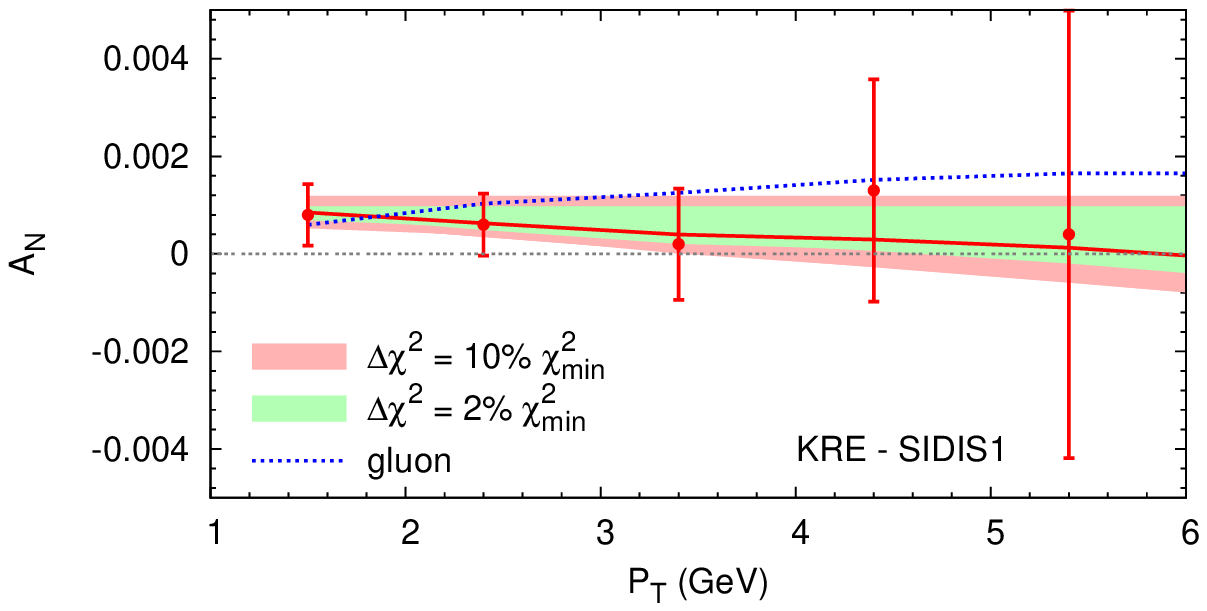}
\caption{Best estimate of the SSA $A_N$, red solid line, compared with PHENIX data~\cite{Adare:2013ekj} at $\sqrt{s}=200$ GeV and at midrapidity, as a function of $P_T$ (in the lower $P_T$ range), obtained adopting the SIDIS2 set~\cite{Anselmino:2008sga} (left panel) and the SIDIS1 set~\cite{Anselmino:2005ea} (right panel) for the quark Sivers functions.
The red(green) band represents a tolerance of 10\%(2\%) in $\chi^2$ (see text for details). The gluon contribution to $A_N$, blue dotted line, is also shown.
 \label{an-lowpt}}
\end{figure}

In Fig.~\ref{an-dss} we present our results for $A_N$ (quark plus gluon contributions) at $\sqrt s = 200$ GeV and midrapidity, compared with PHENIX data~\cite{Adare:2013ekj} and adopting the SIDIS2~\cite{Anselmino:2008sga} extraction of the quark Sivers functions. Here we show the full $P_T$ range, together with our best estimate (solid red line) and a red band corresponding to a tolerance of 10\% in $\chi^2$, as explained above. As one can see the description of data is extremely good, even if the scale adopted in the plot and the tiny data values hide some details. Almost undistinguishable results are obtained for the SIDIS1 set.

To better visualize the data description and the differences between the two sets, in Fig.~\ref{an-lowpt} we show the results for $A_N$ in the lower $P_T$ range for the SIDIS2 (left panel) and the SIDIS1 (right panel) sets. Quite importantly, this is the region that better constrains the gluon Sivers contribution. In this case, we also show the narrower tolerance green band corresponding to a 2\% increase in $\chi^2$, together with the contribution coming from the best estimate of the gluon Sivers function (blue dotted line).

Notice that for the full-$P_T$ range, the Bjorken $x$ explored varies, roughly, between $6\cdot 10^{-3}$ and 0.6, while in the lower-$P_T$ range (up to 5 GeV) the maximum value of $x$ is around 0.4-0.5.
This has to be taken into account, together with the fact that the adopted quark Sivers functions are constrained by available SIDIS data only in the region up to $x \sim 0.3$. In other words, the present analysis, which aims at constraining the gluon Sivers function adopting the information on the quark Sivers contribution and the midrapidity data in $pp$ collisions, is sound only up to $x \sim 0.3-0.4$.
On the other hand, this is the most interesting region for a study of gluon distributions.

\begin{figure}
\includegraphics[width=8cm,angle=0]{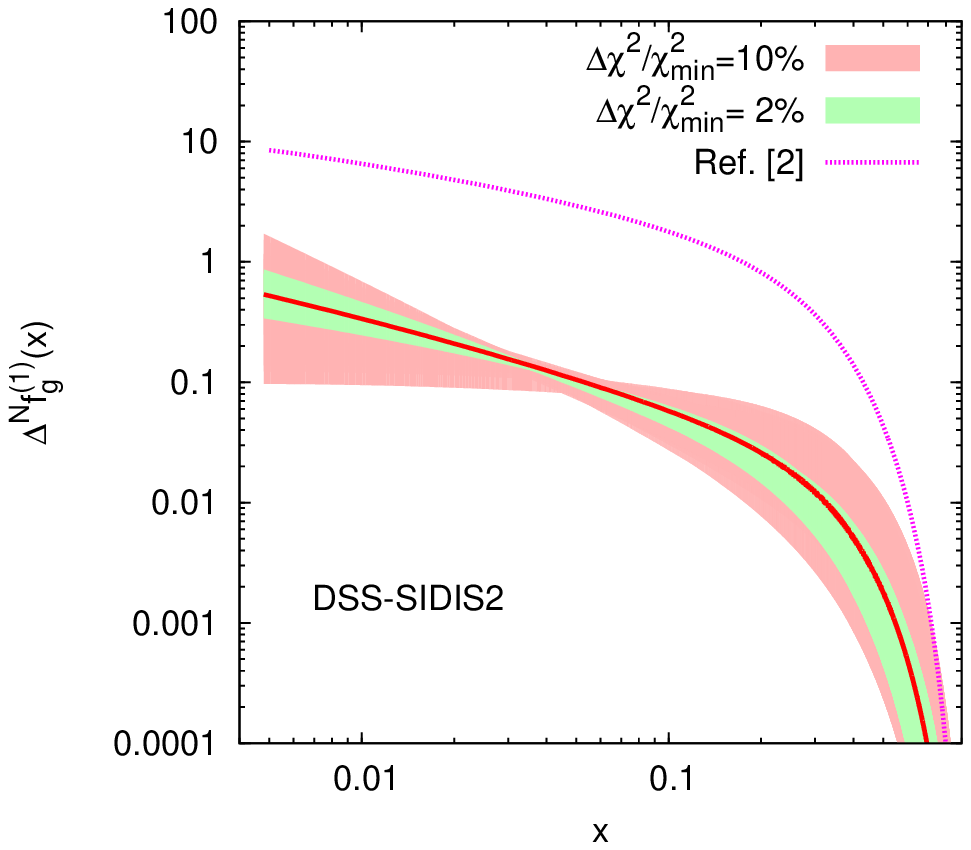}
\includegraphics[width=8cm,angle=0]{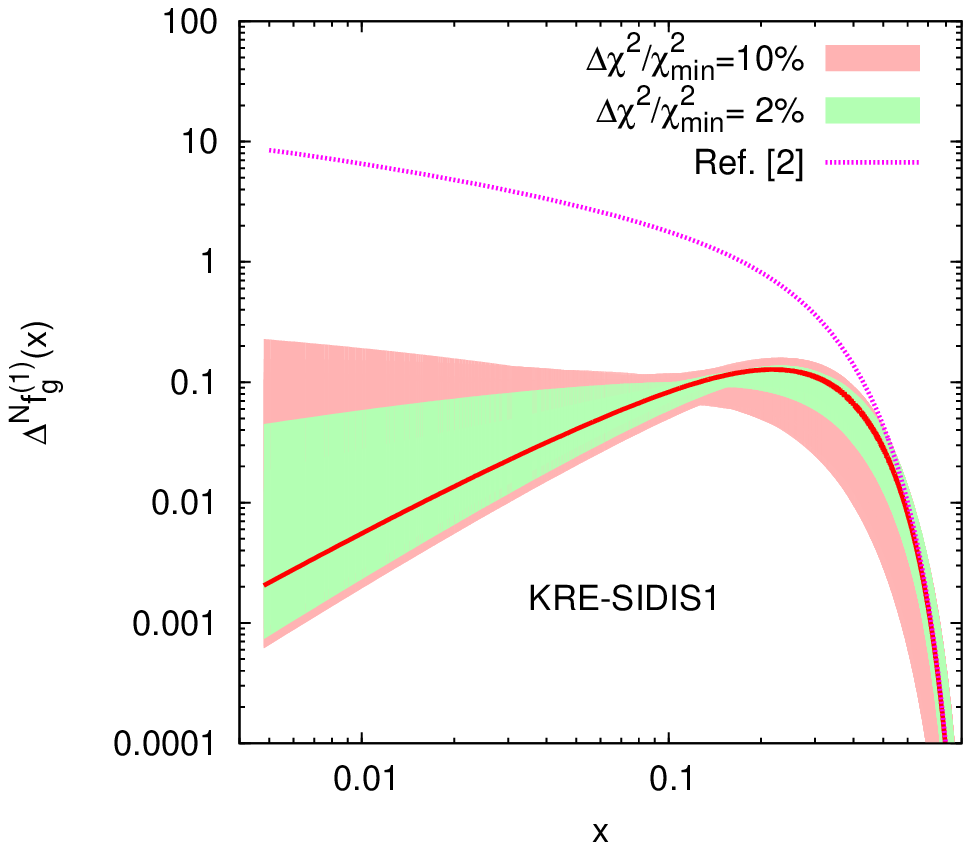}
\caption{First $\bm{k}_\perp$-moment of the gluon Sivers function as defined in Eq.~(\ref{1stmom}) for the SIDIS2 set (left panel) and SIDIS1 set (right panel) at $Q^2=2$ GeV$^2$. The best estimates (red solid lines) are shown together with the tolerance bands corresponding to a 2\% (narrower, green) and 10\% (wider, red) variation in the $\chi^2$. The former bound on the gluon Sivers function (magenta dotted line), obtained in Ref.~\cite{Anselmino:2006yq}, is also shown.
\label{1st-mom}}
\end{figure}

In Fig.~\ref{1st-mom}  we present the corresponding results for the first $\bm{k}_\perp$-moment of the gluon Sivers function, defined as
\be
\label{1stmom}
\Delta^N\! f^{(1)}_{g/p^\uparrow}(x) \equiv \int d^2\bm{k}_\perp \frac{k_\perp}{4M_p}\Delta^N \! f_{g/p^\uparrow}(x,k_\perp) = - f_{1T}^{\perp(1)g}(x)\,.
\ee
More precisely we show (SIDIS2 set in the left panel and SIDIS1 set in the right panel) the best estimates, red solid line, together with the two tolerance bands of 2\% (green, the narrower one) and 10\% (red, the wider one) and the previous upper bound obtained in Ref.~\cite{Anselmino:2006yq} (magenta dotted line). Notice that the two results (old vs.~new bound) for both sets are not directly comparable due to the deep differences in the two analyses. Nevertheless from this new study one can appreciate the tiny role left to the gluon Sivers function when one tries to describe the latest $A_N$ data at midrapidity. This is confirmed even assuming a relatively large tolerance in $\chi^2$, like those considered here.

{}From these results one can quantify the role played by the indeterminacy on the quark Sivers functions and on the fragmentation function sets. This is definitely an important source of uncertainty in the GSF extraction.
In particular, as shown in Fig.~\ref{1st-mom}, we see that the GSF is much smaller (but with larger uncertainties) for the KRE-SIDIS1 case in the low $x$ region, while on the contrary is more constrained for the DSS-SIDIS2 case in the large-$x$ region.
For $0.05 \lesssim x \lesssim 0.3$ the two extractions are almost compatible, considering the uncertainty bands, with the DSS-SIDIS2 bands narrower than the corresponding bands for the KRE-SIDIS1 set in the low-$x$ region, while the viceversa is true in the large-$x$ region. This is related to the fact that the SIDIS2 set has also a more constrained sea quark component and the DSS fragmentation set enhances the role played by the gluon distribution.

Another potential source of uncertainty of this analysis is related to the direct use of the quark Sivers functions as extracted from SIDIS data. As already remarked in the introduction, we do not have a proof of TMD factorization and universality of TMD PDFs for inclusive processes like the one under consideration. Nevertheless, one could speculate about the possible impact of initial and/or final state interactions.

A way to implement these effects in $pp \to \pi\,X$ processes was proposed few years ago in Ref.~\cite{Gamberg:2010tj}, and applied to inclusive pion jet production in $pp$ collisions in Ref.~\cite{D'Alesio:2011mc}, in the framework of the so-called color gauge-invariant (CGI) GPM approach. We recall here that the authors of these works focused only on the quark initiated processes and that nothing has been done so far on the gluon sector. To account also for this source of uncertainty we have reconsidered the contribution of the quark Sivers functions adopting the CGI-GPM. It is important to note that differently from what happens in the forward rapidity region, where one gets a contribution of almost the same size but opposite in sign w.r.t.~the GPM, in the midrapidity region the overall effect is a strong reduction in size, but keeping the same sign. This is due to the fact that in this kinematical region many partonic channels play a comparable role, leading to relative cancellations among their contributions. The use of this result in the present analysis would imply a reduction of the GSF and a relative larger indeterminacy towards its smaller values.

In the spirit of further pursuing this issue we explore a somewhat more extreme scenario, maybe less realistic, but worth of being considered. We repeat the procedure described above, adopting a quark Sivers function reversed in sign w.r.t.~the one extracted from SIDIS, even if we are aware that for the process under consideration one would expect a more involved structure. We think that this attempt should give a clear indication of the most extreme variation of the GSF uncertainty.

\begin{figure}
\includegraphics[width=8cm,angle=0]{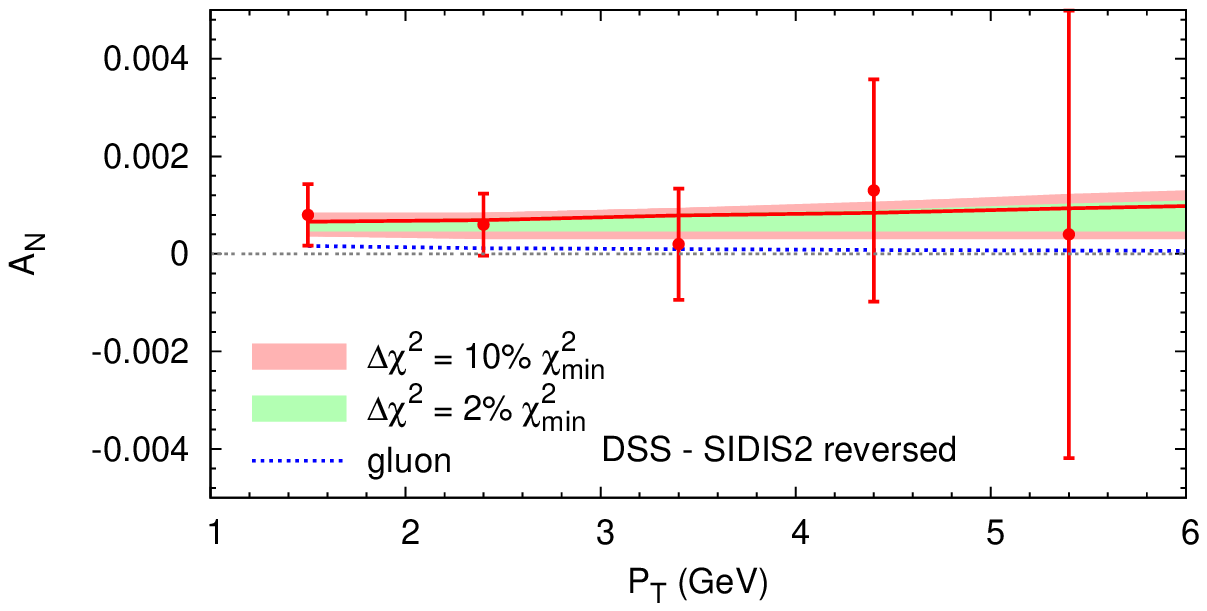}
\includegraphics[width=8cm,angle=0]{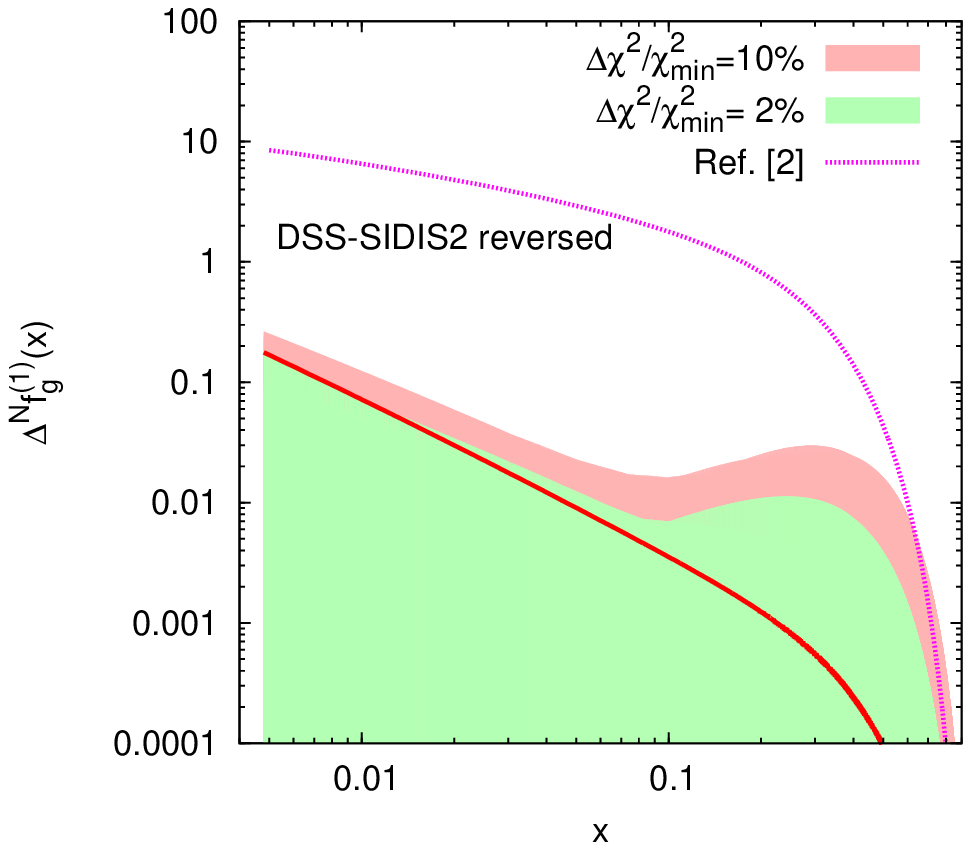}
\caption{Left panel: Best estimate of $A_N$, red solid line, compared with PHENIX data~\cite{Adare:2013ekj} at $\sqrt{s}=200$ GeV and at midrapidity, as a function of $P_T$ (in the lower $P_T$ range), obtained adopting the SIDIS2 set~\cite{Anselmino:2008sga}  for the quark Sivers functions reversed in sign.
The red(green) band represents a tolerance of 10\%(2\%) in $\chi^2$ (see text for details). The gluon contribution to $A_N$, blue dotted line, is also shown.
Right panel: First $\bm{k}_\perp$-moment of the gluon Sivers function as defined in Eq.~(\ref{1stmom}) for the SIDIS2 set (with the quark parameterizations reversed in sign) at $Q^2=2$ GeV$^2$. The best estimate (red solid line) is shown together with the tolerance bands corresponding to a 2\% (narrower, green) and 10\% (wider, red) variation in the $\chi^2$. The former bound on the gluon Sivers function (magenta dotted line), obtained in Ref.~\cite{Anselmino:2006yq}, is also shown.
\label{AN-1st-mom-reversed}}
\end{figure}

In Fig.~\ref{AN-1st-mom-reversed} we show the results for the SIDIS2 case, that corresponds to the most striking effect. In the left panel, one can easily see that the description of $A_N$ is now given in terms of the quark Sivers function alone and that the GSF contribution is almost negligible. This reflects also in the first moment of the GSF, right panel. Here, beside its stronger suppression (red solid curve) compared to the GPM result (left panel of Fig.~\ref{1st-mom}), the uncertainty bands, even for a tolerance of 2\%, extend to very low values, compatible with zero.
Analogous considerations concerning the stronger suppression of the GSF in the effectively explored region apply also to the SIDIS1 case.

It is evident that such an extreme scenario would imply an even more negligible role of the GSF, that could result almost compatible with zero.

In summary, and with all the cautions discussed above, in the $x$ region explored by $pp$ data the gluon Sivers function can be effectively constrained. It results to be positive, strongly suppressed with respect to the previous bound~\cite{Anselmino:2006yq} and much smaller than its positivity bound.

\section{Conclusions}

In this paper we have analyzed the impact of recent, highly precise, data for the transverse single spin asymmetry
$A_N(p^\uparrow p\to\pi^0\,X)$ at central rapidity and moderately large transverse momentum measured by
the PHENIX Collaboration at RHIC on our knowledge of the gluon Sivers function.

To this aim we have utilized the so-called transverse momentum dependent generalized parton model which takes into account intrinsic parton motion and spin effects, extending the well-known collinear leading order parton model.

Adopting the most recent phenomenological information, within the same approach, on the (sea) quark Sivers distributions, coming from SIDIS data, we have shown how the PHENIX data allow us to constrain the GSF considerably, as compared to the positivity bound as well as to a previous bound based uniquely on less precise $A_N$ data.

We have found that the new constraint is particularly significant, within theoretical uncertainties, in the region of gluon momentum fraction $0.05 \lesssim x \lesssim 0.3$, that is the presently explored SIDIS region, where the quark Sivers distributions are well constrained. At lower-$x$ values the bound is still effective but theoretical uncertainties become large, as expected. At larger-$x$ values the bound is looser; on the other hand this is the region where the relevance of gluon contributions is small due to dominance of quark channels.

We have also considered midrapidity data measured by the STAR collaboration for $pp\to {\rm jet}\, X$ processes~\cite{Adamczyk:2012qj}, where one can access directly the TMD-PDFs. On the other hand these data do not improve the constraint on the GSF due to the $P_T$ region explored and their relatively large error values. We have nevertheless checked that the new bound is consistent with these data.

We can then say that the present analysis, constraining the GSF in size and sign (a new aspect w.r.t.~Ref.~\cite{Anselmino:2006yq}), strongly reduces the possible role of the GSF in spin and azimuthal asymmetries for processes covering $x$ regions similar to those explored here.

Some words of caution are however required: factorization has not been proven in the context of the TMD-GPM for single inclusive processes in proton-proton collisions. In our approach TMDs keep their partonic interpretation and universality.
However, initial and/or final state interactions, required to preserve color gauge invariance, might spoil the factorization for these processes, leading to process dependence and universality breaking effects.
It is not easy to figure out only from theoretical considerations the phenomenological relevance of these possible
effects for currently accessible processes.
As a matter of fact, nowadays the TMD-GPM model is able to reproduce fairly well, within uncertainties, the majority of experimental data available on azimuthal and single spin asymmetries in SIDIS and proton-proton collision processes.
Moreover, from presently available data there is no clear and unambiguous evidence of sizable universality breaking effects.

However, to investigate, even in an approximate way, the potential role of the process dependence of the quark Sivers functions, we have also considered an extreme scenario, adopting the quark Sivers function as extracted from SIDIS but reversed in sign. In such a case in the explored region one would get an even more suppressed GSF, with uncertainty bands extending to values compatible with zero.

For these reasons, we believe that, even if with some caution, the constraints on the GSF resulting from this analysis are sound within uncertainties and must be taken into account in further phenomenological analyses involving such TMD distribution.

Less involved processes from the point of view of color gauge links, where proving factorization might be easier, need to be considered to further clarify the issues concerning process dependence and test the smallness of the GSF.  For example, processes like $e p^\uparrow \to Q \bar{Q}\,X$, or $e p^\uparrow \to {\rm jet}\, {\rm jet}\,X$, where only final state interactions (like in SIDIS) are involved, could be studied at future electron-ion
colliders (EIC) \cite{Boer:2011fh}. Analogously, in $p^\uparrow p\to \gamma\gamma\,X$ \cite{Qiu:2011ai} or $p^\uparrow p\to J/\psi\; \gamma\,X$ processes only initial state interactions are involved like in DY.
Moreover, factorization has been already proved, at the next-to-leading order, for $p^\uparrow p\to \eta_{c,b}\,X$ \cite{Ma:2012hh} and $A_N$ for this process might be measurable at the proposed AFTER@LHC set-up \cite{Brodsky:2012vg, Massacrier:2015nsm}.

In Refs.~\cite{Gamberg:2010tj, D'Alesio:2011mc} a first attempt to compute initial and final state interactions for SSAs in hadronic collisions within the TMD-GPM approach, focusing on quark initiated subprocesses, and studying their phenomenological consequences, has been made. 
It would be very interesting to extend this kind of analysis to gluon initiated subprocesses, relevant to the present case, and study its effects on the bounds for the gluon Sivers function.

\acknowledgments
We would like to thank K.~Barish and J.~Koster for information on the data, and M.~Anselmino and S.~Melis for their critical reading of the manuscript.
We acknowledge support of the European Community under the FP7 ``Capacities - Research Infrastructures'' program (HadronPhysics3, Grant Agreement 283286). U.D.~is grateful to the Department of Theoretical Physics II of the Universidad Complutense of Madrid where part of this work was completed. C.P.~acknowledges support by the “Fonds Wetenschappelijk Onderzoek - Vlaanderen” (FWO) through a postdoctoral Pegasus Marie Curie Fellowship.


\end{document}